\documentstyle[aps,preprint,epsfig]{revtex}

\begin{document}

\preprint{{\large SUNY-NTG-97-2}}

\title{Nucleon flow and dilepton production in heavy-ion collisions}
\bigskip
\author{G. Q. Li$^{1,2}$, G. E. Brown$^1$, C.-H. Lee$^1$, and C. M. Ko$^2$}
\address{$^{1}$ Department of Physics, State University of New
York at Stony Brook, Stony Brook, NY 11794\\
$^{2}$ Cyclotron Institute and Physics Department,
Texas A\&M University, College Station, Texas 77843}
 
\maketitle
 
\begin{abstract}
Nucleon flow in Au+Au collisions at 1 GeV/nucleon region and
dilepton spectra in S+Au collisions at SPS energies are
analysed in the relativistic transport model. They are shown 
to be sensitive to the nucleon vector and scalar potentials, 
respectively, and are thus useful probes of these potentials
in dense matter. A recently developed effective Lagrangian with
self-interactions of vector as well as scalar fields is found to
describe both data reasonably well. 
\end{abstract}

\pacs{25.75.+r, 24.10.Jv}
 
One of the primary motivations of high-energy heavy-ion collisions
is to study the properties of hadrons in dense matter, which are
expected to reflect
the chiral symmetry breaking in vacuum and its restoration
at high temperature and/or density, the so-called chiral phase
transition \cite{br96a}. Closely related to this is the study of 
deconfinement, namely, the formation of quark-gluon 
plasma in heavy-ion collisions \cite{bh96}. The properties
of dense matter that can be extracted from heavy-ion collisions
are also very useful for astrophysics problems, such as the properties
of neutron stars \cite{prak95}. There are basically two types of
observables that can be consided as `clean' probes of 
hadronic properties in dense matter. One of these are the
various collective `flows' of hadrons that are sensitive to
the entire dynamical process of heavy-ion collisions, and the other
is the electromagnetic observables such as photon and 
dilepton spectra that are not affected by strong final-state
interactions and thus carry information about the early 
stage of heavy-ion collisions.

Collective flows of nucleons and light fragments in heavy-ion collisions
have been used to study the momentum as well as density
(related to equation of state of equilibrium nuclear matter) 
dependence of the nucleon potential \cite{gut89,eos,ags,panzhang}.
In Boltzmann-Uehling-Uhlenbeck (BUU) calculations, it has been
found that a soft equation of state with a momentum dependent 
nucleon potential describes well the experimental data \cite{panzhang}.
In relativistic transport models \cite{ko87,mos93} based on various
Walecka-type $\sigma$-$\omega$ models \cite{qhd}, 
the corresponding Schr\"odinger equivalent potential is explicitly
momentum dependent. In the mean-field approximation,  it
increases linearly with nucleon kinetic energy \cite{maru94}. In order
to describe experimental data on flow in the 1 GeV/nucleon region
in the mean field approximation, 
it was found that one has to use a relatively weak vector potential 
with $g_\omega^2/4\pi \approx 4.0$, and consequently also a weak 
scalar potential \cite{liko95a}. The magnitudes of these potentials
are much smaller than those in the Walecka linear $\sigma$-$\omega$ model
(hereafter referred to as Walecka model) \cite{qhd}. 

Photon and dilepton spectra have also been studied extensively,
both experimentally and theoretically, especially for heavy-ion
collisions at SPS energies. A particular interesting observation 
from these experiments is the enhancememt of low-mass dileptons 
in heavy-ion collisions as compared to proton-induced reactions
\cite{ceres,helios}. Theoretical calculations from
various groups with different dynamical models converge to one conclusion,
namely, medium effects are needed to explain the data \cite{lkbs96,others}.
Based on the initial conditions provided by the relativistic 
quantum molecular dynamics (RQMD) \cite{rqmd}, 
we found that the data can be well accounted if the rho meson mass 
decreases at the same rate as the nucleon effective mass, i.e. \cite{br91}, 
\begin{eqnarray}
{m_\rho ^*\over m_\rho}\approx {m_\omega ^*\over m_\omega}
\approx {m_N^*\over m_N},
\end{eqnarray}
with the nucleon effective mass obtained 
from the Walecka model \cite{lkbs96}. 

There are, however, concerns about the use of Walecka model.
This model does not describe nuclear matter saturation
properties well; especially a compression modulus of
about 560 MeV is simply too large as compared with the empirical value
of about 200 MeV. The use of the Walecka model for heavy-ion 
collisions in the 1 GeV/nucleon region would overpredict the nucleon
flow data, even after the incorrect momentum dependence in the 
mean-field potential is properly taken care of, as we shall discuss later.

The main purpose of this Letter is to show that there 
are effective chiral Lagrangians that can describe consistently 
{\it both} the flow and dilepton data, {\it as well as} nuclear
matter and finite nuclei properties. We will further show that
the nucleon flow is chiefly sensitive to the vector potential,
while the shape of dilepton spectra is mainly 
determined by the scalar potential. They can thus be 
individually used to  probe these potentials at higher densities. 
This separation is important for studies of hyperon and
kaon properties in dense matter.

The effective Lagragian we use in this study is the one recently
developed by Furnstahl, Tang, and Serot based on chiral 
symmetry considerations and the naturalness argument 
\cite{fst95,fst96}. They introduced, in addition to the 
self-interaction of the scalar field, the self-interaction of
the vector field and coupling between the scalar and the vector
field. In their energy density functional formalism,
the energy density for the symmetric nuclear matter 
is given by \cite{fst96}
\begin{eqnarray}
{\cal E}(\Phi, W; \rho _B) = W \rho _B -{1\over 2C_V^2} W^2 +{\bar \alpha}
m_N \Phi W^2 -{\zeta \over 24} W^4 \nonumber\\
+ {1\over 2C_S^2} \Phi ^2
+{{\bar \kappa}\over 6} \Phi^3 +{{\bar \lambda} \over 24} \Phi^4
+{4\over (2\pi )^3}\int d^3p \sqrt {{\bf p}^2+m_N^{*2}},
\end{eqnarray}
where $m_N^*=m_N-\Phi$ and $\Phi$ and $W$ are the scalar and vector
fields, which couple to nucleon with coupling constants
$C_S$ and $C_V$, respectively. The scalar field shifts the
nucleon mass, and the vector field its energy. If $\zeta$ and
${\bar \alpha}$ are zero, this reduces to the usual non-linear 
$\sigma$-$\omega$ model \cite{liko94}. If furthermore ${\bar \kappa}$
and ${\bar \lambda}$ are set to zero, we recover the original
linear Walecka model \cite{qhd}. The introduction of 
the vector field self-interaction and the coupling between
the vector and scalar fields are useful for making a relatively soft
equation of state, and at the same time producing a small nucleon
effective mass. 

By fitting to nuclear matter as well as finite nuclei properties,
Furnstahl {\it et al.}\cite{fst95,fst96} proposed
a number of parameter sets. Here we will use one of these as listed
in the first row of Table 4 of Ref. \cite{fst96}, which we denote as 
the FTS-T1 model. This parameter set leads to a saturation density 
of about 0.15 fm$^{-3}$, with a binding energy of about 16 MeV. 
The nucleon effective mass $M_N^*/M_N$ and the compression modulus 
at nuclear matter density are about 0.6 and 200 MeV, repectively.
We show in Fig. 1(a) the nucleon effective mass as a function of 
density. The dotted curve shows the nucleon effective
mass in the Walecka model \cite{qhd}. Up to 3$\rho_0$ 
they are very similar to each other.

\begin{figure}
\epsfig{file=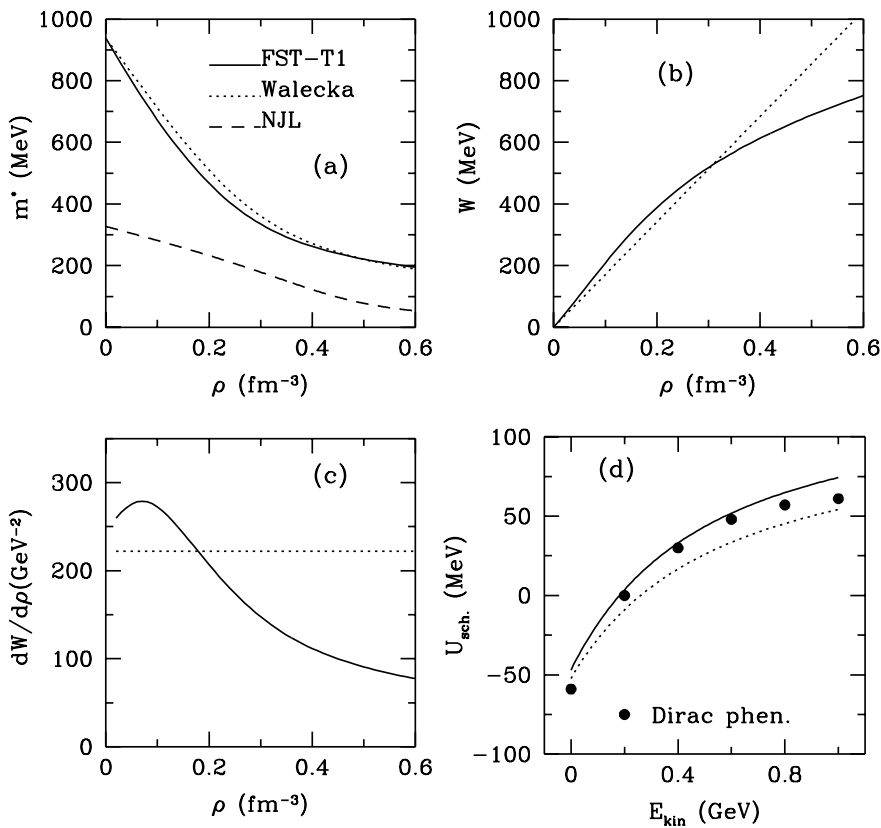,height=5.0in,width=5.0in}
\caption{(a) Effective mass of nucleon (soild and dotted curve)
and constituent quark mass (dashed curve) as calculated
in the NJL model with bare quark mass of 5 MeV, (b) vector potential,
(c) density derivative of vector potential,  and (d)
Schr\"odinger equivalent potential in the FST-T1 
model and the Walecka model.}
\end{figure}

Also of interest is the comparison of the vector potential with and 
without the vector field self-interaction and vector-scalar coupling.
When $\zeta$ and ${\bar \alpha}$ are zero, the vector potential
is simply given by $C_V\rho_B$, so it rises linearly with density
as in the Walecka model. The vector potential in Walecka model
($g_\omega ^2/4\pi\approx 10.6$) is shown in Fig. 1(b)
by the dotted curve, and compared with that from the
FST-T1 model ($g_\omega ^2/4\pi\approx 12.0$).
At low densities where self-interaction terms are unimportant, the
vector potential in the FST-T1 model is larger than that in
the Walecka model, because of a larger vector coupling constant in
the former. However, as density increases, the high-order terms
become important, and reduce the vector potential in the FST-T1
model. Around 3$\rho _0$ the vector potential reduces from about 810 MeV
in the Walecka model to about 670 MeV in the FST-T1 model.
We will show that this bending down of the vector potential is very 
important to reproduce correctly the nucleon flow data,
as it also reduces the slope, $dW/d\rho$, that enters
the Hamilton equation of motion in the transport model 
which changes nucleon momentum. This is shown in Fig. 1(c)
and compared with that in the Walecka model which is just the 
coupling constant $C_V$. This effect also
softens the nuclear equation of state and helps to reduce the
maximum mass of neutron stars \cite{serot96}. On the
microscopic level, the Dirac-Brueckner-Hartree-Fock (DBHF)
calculation also shows the nonlinear density dependence of
the vector potential \cite{toki}.

On the mean-field level, the scalar and vector potentials
are dependent only on density but not on momentum. Higher-order
contributions introduce momentum depenence. 
In the DBHF calculation of ter Haar and Malfliet \cite{mal87},
this momentum dependence was studied, and the scalar and vector
potentials were found to decrease with momentum. 
Empirical information about this comes mainly from the
Dirac phenomenology analysis of the intermediate-energy
proton-nucleus scattering \cite{clark}. 
In the present study we use a schematic approach to
take account of this momentum dependence. 
At each local density, we solve the FST-T1 model
in the mean-field approximation and obtain nucleon scalar and vector
potentials that are momentum independent. The effective potentials 
for a nucleon with kinetic energy $E_{kin}$ (relative to its 
local cell) are then obtained by scaling the momentum independent 
potentials with the momentum dependence extracted from the 
Dirac phenomenology \cite{clark}. The Schr\"odinger equivalent
potential obtained in this way are shown in Fig. 1(d) and found to be
in good agreement with Dirac phenomenology for both the FST-T1 and 
the Walecka model. 

We note that this treatment of the momentum dependence is not fully 
self-consistent. Since the introduction of the momentum dependence
changes the equilibrium nuclear matter properties, a
self-consistent procedure is required to satisfy both the 
equilibrium and non-equilibrium properties. This amounts to
a small change of the parameters in the original model and
also some small changes in the fitting of Dirac phenomenology. 
Since the energy scale involved in equilibrium nuclear matter
problem is much smaller than that in the initial stage of
high-energy heavy-ion collisions, the effect of this self-consistent
procedure on our flow results is expected to be unimportant.

Our results for nucleon flow parameter $F$ in Au+Au collisions
in the 1 GeV/nucleon region are shown in Fig. 2 together with
the experimental data from the EOS Collaboration \cite{eos}.
The cascade calculation which does not include any mean field 
potential accounts already for about 2/3 of the observed
flow. This is consistent with the results of BUU model of Ref.
\cite{panzhang}. In ARC calculations\cite{arc}, however, nucleon
flow turns to be somewhat stronger and is thus close to the data, 
since an effectively repulsive trajectory is used for two-body scattering.

\begin{figure}
\epsfig{file=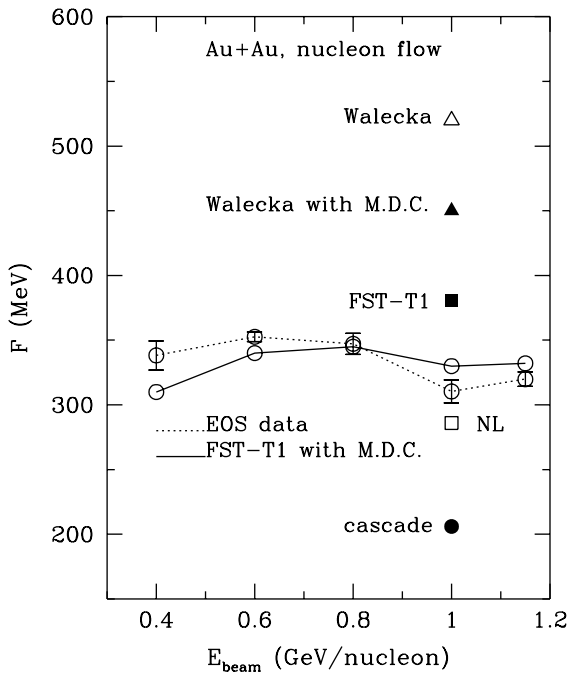,height=4.0in,width=4.0in}
\caption{nucleon flow parameter in Au+Au collisions 
based on various effective Lagrangians.}
\end{figure}

The result of Ref. \cite{liko95a} using the non-linear $\sigma$-$\omega$
model is shown in Fig. 2 by the open square. The flow parameter
is about 280 MeV and is in good agreement with the data.
Although no energy correction was introduced in Ref. \cite{liko95a},
because of the use of relatively weak scalar and vector potentials,
the Schr\"odinger equivalent potential in the non-linear $\sigma$-$\omega$
model turns out to be in reasonable agreement with the Dirac phenomenology.

As mentioned, the Walecka model produces too much flow. 
The nucleon flow parameter obtained with the Walecka model without the
momentum dependence correction (M.D.C.) is shown in Fig. 2 by 
the open triangle, while that with the M.D.C. is shown by solid triangle.
The flow parameter in the first case is about 520 MeV, and is much too
large as compared with the data. After including the M.D.C.,
which reduces the repulsion in the initial stage, 
the flow parameter reduces to about 450 MeV, which is still
too large as compared to the data.

The result of the FST-T1 model without the M.D.C.
is shown in the figure by the solid square, and is about 380 MeV.
This is considerably smaller than that of the  Walecka model 
with and without the M.D.C., but somewhat larger than the
data. The results of the FST-T1 model with the M.D.C.
are shown in Fig. 2 by open circles connected by solid lines.
At 1 GeV/nucleon, the flow parameter reduces to about 330 MeV and
is in now good agreement with the data, as is the case at other
beam energies. We can thus conclude that with a proper correction 
for the momentum dependence, the FST-T1 model can explain the nucleon 
flow in heavy-ion collisions at 1 GeV/nucleon region quite well.

At $\rho _0$ and below, the scalar and the vector potentials, as well as
the corresponding Schr\"odinger potentials in the FST-T1 model
and Walecka model are very much the same (see Fig. 1). 
Actually, the scalar potentials in the two models agree with each other
up to 3$\rho_0$. Then why do they
generate substantially different nucleon flow? The reason lies
in the difference in their vector potentials at high densities.
Because of high-order terms involving vector field self-interaction
and vector-scalar coupling, the vector potential in the FST-T1 model
becomes smaller than that in the Walecka model at high densities, 
thus providing less repulsion. Also the 
slope of the vector potential, $dW/d\rho $, gets smaller 
at high densities, providing less repulsive force.
The nucleon flow is indeed sensitive to nucleon vector potential
at high densities (2-3$\rho_0$ in Au+Au collisions). 

In Fig. 3 the dilepton spectra in central S+Au collisions at 
200 GeV/nucleon obtained in the FST-T1 model is shown. 
The calculations were performed as in Ref. \cite{lkbs96} but
with the Walecka model replaced by the FST-T1 model. The vector
meson masses were assumed to scale like the nucleon mass,
as in Eq. (1). It is seen that the fit with the in-medium meson
masses is acceptable, although somewhat lower than the
center of the experimental points in the low-mass region,. Given
the similarity between scalar fields in the FST-T1 and the
Walecka model, the similarity in the dilepton
production in not surprising.

\begin{figure}
\epsfig{file=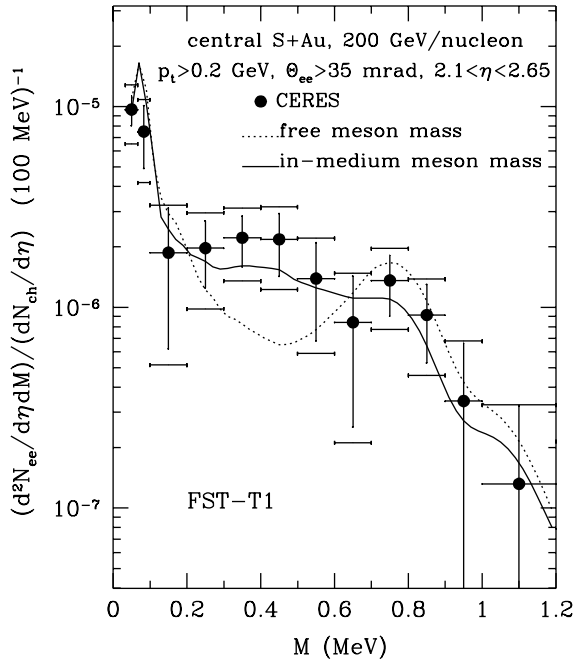,height=4.0in,width=4.0in}
\caption{Dilepton spectra in central S+Au collisions at 200 
GeV/nucleon with free (dotted curve) and in-medium (solid curve)
meson masses. The experimental data from Ref. [13] are
shown by solid circles, with systematic errors given by bars.
Brackets represent the square root of the 
quadratic sum of the systematic and statistical errors.}
\end{figure}

Note that the nucleon flow (coming from the Vlasov part of the transport
model) depends chiefly on the vector interaction. 
The force acting on a nucleon is proportional to the
spatial derivative of its energy, $({\bf p}^2+m_N^{*2})^{1/2}+W$. 
In the region of densities important for flow (2-3$\rho_0$), the nucleons
behave essentially as massless particles of momentum $p$ in a 
vector potential $W$, since $m_N^*$ is small compared to $p$ 
On the other hands, the shape of the dilepton spectra 
depends almost completely on the scalar field, since the effective mass
of vector mesons ($\rho ,~\omega$), and hence the dilepton mass,
are related to the nucleon scalar field through Eq. (1). From the 
fitting to the CERES data, it is found that the scalar field
must be sufficiently strong to reduce significantly 
the nucleon effective mass; and through Eq. (1), 
the effective mass of vector mesons.

We should remark that the vector mean field depends on the
ratio of $g_{\omega NN}^*/m_\omega^*$. If $m_\omega^*$ decreases
in medium like $m_\rho^*$, then $g_{\omega NN}^*$ must also
decrease so that the vector potential increases at most linearly
with density. The vector coupling is expected to scale in theories 
which treat the vector mesons as gauge particles of
the hidden symmetry \cite{hara93}. Unfortunately, it is difficult
to calculate quantitatively in the nonperturbative regime discussed
here. However, the nucleon flow does empirically determine  
the ratio $g_{\omega NN}^*/m_\omega^*$ to be independent of density
at low densities and to decrease slightly at high densities,
 
Recently, Brown, Buballa, and Rho \cite{bbr96} argue
for a chiral restoration transition that is mean field
in the NJL model. In this case the 
scalar coupling $g_{\sigma NN}$ should remain unchanged 
through the transition. Brown {\it et al} find that
the transition takes place at $\rho_c \le 3\rho _0$.
The NJL model has dropping scalar meson mass, because 
$m_\sigma^* = 2 m_Q^*$, where $m_Q^*$ scales with density.
Once one introduces a bare quark mass of about 5 MeV, however, $m_Q^*$
does not drop so rapidly, as shown in Fig. 1(a). 
At 3$\rho_0$, $m_Q^*\approx 70$ MeV, which is just about
1/3 of the $m_N^*$ at that density in the FST-T1 model.
The vector field in NJL does not participate in the
phase transition, and must be put in ``by hand''. Thus we can say
that the constituent quark mass in the NJL model, once bare quark
mass are included, is about the same in the the density
region $\rho \approx 3\rho_0$, as in the FST-T1 model. 
Since dileption calculation depends chiefly on the scalar field,
it seems clear that the NJL model will produce about the same
dileptons as the FST-T1 model. This is an important statement, 
because the NJL model has the symmetries of QCD.
The Walecka-type models can be derived \cite{gel95,br96}
so that they possess chiral symmetry in low order, but do not
have relevant higher order chiral terms contained in
the NJL model.

In summary, we used the effective Lagrangian recently developed
by Furnstahl, Tang and Serot in the transport model analysis
of nucleon flow and dilepton production in heavy-ion 
collisions.  We found that the model describe both sets of
data reasonably well. The ability to describe the dilepton
spectra can be attributed to the strong scalar field
of the model that reduce significantly the vector meson
mass through Eq. (1). The ability to describe the nucleon flow data
can be traced back to the inclusion of vector field 
self interactions that reduce the vector potential at
higher densities. 
 
We would like to thank M. Prakash and Mannque Rho for many helpful
discussions. This work was supported in part by the Department
of Energy under Grant No. DE-FG02-88ER40388, and by the 
National Science Foundation under Grant No. PHY-9509266.  
The work of CHL was partly supported by Korea Science and
Engineering Foundation.

\end{document}